\documentclass[lettersize,journal]{IEEEtran}
\usepackage{amsmath,amsfonts}
\usepackage{array}
\usepackage[caption=false,font=normalsize,labelfont=sf,textfont=sf]{subfig}
\usepackage{textcomp}
\usepackage{stfloats}
\usepackage{url}
\usepackage{verbatim}
\usepackage{graphicx}
\usepackage{cite}
\usepackage{booktabs}
\usepackage{multirow}
\usepackage{pifont}
\newcommand{\cmark}{\ding{51}}
\newcommand{\xmark}{\ding{55}}

\begin{document}

\title{A Human-Like Pedestrian Model for Automated Driving Simulations}

\author{Ruofeng~Wang, Patrick~Ebel, Philipp~Wintersberger, and Antti~Oulasvirta%
\thanks{Ruofeng Wang and Antti Oulasvirta are with the Department of Information and Communications Engineering, Aalto University, Espoo, Finland.}%
\thanks{Patrick Ebel is with the Hasso Plattner Institute, University of Potsdam, Potsdam, Germany.}%
\thanks{Philipp Wintersberger is with Interdisciplinary Transformation University Austria, Austria.}%
\thanks{This work has been submitted to the IEEE for possible publication. Copyright may be transferred without notice, after which this version may no longer be accessible.}
\thanks{Corresponding author: Antti Oulasvirta (e-mail: antti.oulasvirta@aalto.fi).}%
}

\maketitle

\begin{abstract}

Automated vehicles must be able to interact with pedestrians safely and efficiently across diverse traffic situations. 
Although driving simulators offer a scalable testbed for learning such capabilities, existing theory-inspired pedestrian models are narrow in scope and limited to go/no-go crossing decisions in single-lane settings. While data-driven approaches can predict pedestrian behavior in complex situations, they lack sufficient observations in rare, safety-critical scenarios.
Here, we propose an approach to training pedestrian models in simulators so that learned policies generate demonstrably human-like behavior in realistic, complex traffic scenarios, including multiple lanes, heavy traffic, and dangerous driving styles.
Our technical contribution is a novel definition of pedestrian--vehicle interaction as a partially observable Markov decision process (POMDP) with theory-grounded perceptual, cognitive, and motor constraints. 
It accounts for the highly adaptive nature of human behavior in traffic and simulates how people adjust their responses according to perceived danger, time pressure, and the complexity of the situation.
When trained via deep reinforcement learning (RL) with domain randomization in a simulator, the model reproduces the broadest range of empirical findings shown so far on human crossing behavior, including gap acceptance, yielding acceptance, hesitation, and evasive speed adjustment. 
We show that learned policies transfer to unseen traffic environments, and can be further adapted to local traffic norms with finetuning. 
Together, these results establish a blueprint for simulator-ready pedestrian models that can support the development and evaluation of automated driving systems.
\end{abstract}

\section{Introduction}
% Introduces the objective and motivates it
Modeling how pedestrians behave in traffic is an important aspect of developing safe and efficient automated vehicles and advanced driver-assistance systems~\cite{arellana2022analyzing,jayaraman2020efficient,herman2022pedestrian}.
While the capabilities for detecting and tracking pedestrians have improved, it is an open problem how to plan actions while factoring in probable human responses.
Accordingly, high-fidelity models of pedestrian behavior are needed to support the training, testing, and evaluation of automated vehicles. 
Developing such models remains challenging. 

Drawing on prior work on pedestrian modeling and pedestrian--vehicle interaction~\cite{Camara2021,rasouli2019autonomous}, we consider three practical requirements for pedestrian models intended for automated-driving simulation. First, they should reproduce empirically documented behavior, from road crossing initiation~\cite{petzoldt2014relationship,tian2023deceleration}, through adaptive responses after entering the roadway~\cite{balwan2017impacts,zhuang2011pedestrians}, to the selection of crossing strategies in multi-lane traffic~\cite{zhang2019evaluation}. Second, the resulting pedestrian behavior should generalize across traffic conditions, including variations in lane count, vehicle speed, and traffic density~\cite{tian2022explaining,theofilatos2021cross}, while remaining adaptable to local traffic norms and conventions~\cite{pele2017cultural}. Third, the models should be deployable as closed-loop agents in traffic simulators~\cite{dosovitskiy2017carla,rempe2023trace,yang2023suicidal}.

% Research gap
However, existing pedestrian crossing models fall short of this goal. We divide current approaches into three families based on their purpose: explanation, prediction, and use in simulation. 
Explanatory models offer theoretical accounts of how pedestrians decide to cross the street, such as drift diffusion models that capture how perceptual uncertainty accumulates over time before a crossing decision is made~\cite{pekkanen2022variable,ratcliff2016diffusion}. 
Binary discrete-choice and gap-acceptance models treat crossing as a single go/no-go decision governed by traffic-context variables~\cite{yannis2013pedestrian,tian2022explaining,amini2021towards,opiew2023study}. 
If predicting pedestrian intention and trajectories is the main purpose, deep learning models that leverage large-scale video or multimodal data~\cite{fang2018pedestrian,rasouli2019pie,rasouli2017they,razali2021pedestrian,yang2022predicting,wang2024multi,yang2024hierarchical} are the preferred choice. 
Simulation-oriented models mostly rely on rule-based or controller-based designs to generate executable pedestrian behavior, including CARLA’s AI walker, ``suicidal'' pedestrian models and pedestrian trajectory generation methods~\cite{dosovitskiy2017carla,priisalu2022generating,yang2023suicidal,lu2025can,rempe2023trace}. 

Although all three families address important aspects of pedestrian behavior, few individual models combine broad empirical behavioral coverage with generalization and closed-loop simulator deployment. Developing a unified model with these capabilities therefore remains an important open problem~\cite{zhang2023pedestrian,gesnouin2022assessing,galvao2024pedestrian,rasouli2019autonomous}.

This gap significantly impacts the development of automated vehicles. Without simulator-ready pedestrian models that accurately reflect how crossing behavior changes based on vehicle dynamics, traffic density, and crosswalk design, automated systems trained in simulations experience distribution shifts when they encounter real pedestrians. This can lead to dangerously inappropriate responses in critical situations.

\begin{figure*}[!t]
    \centering
    \includegraphics[width=\textwidth]{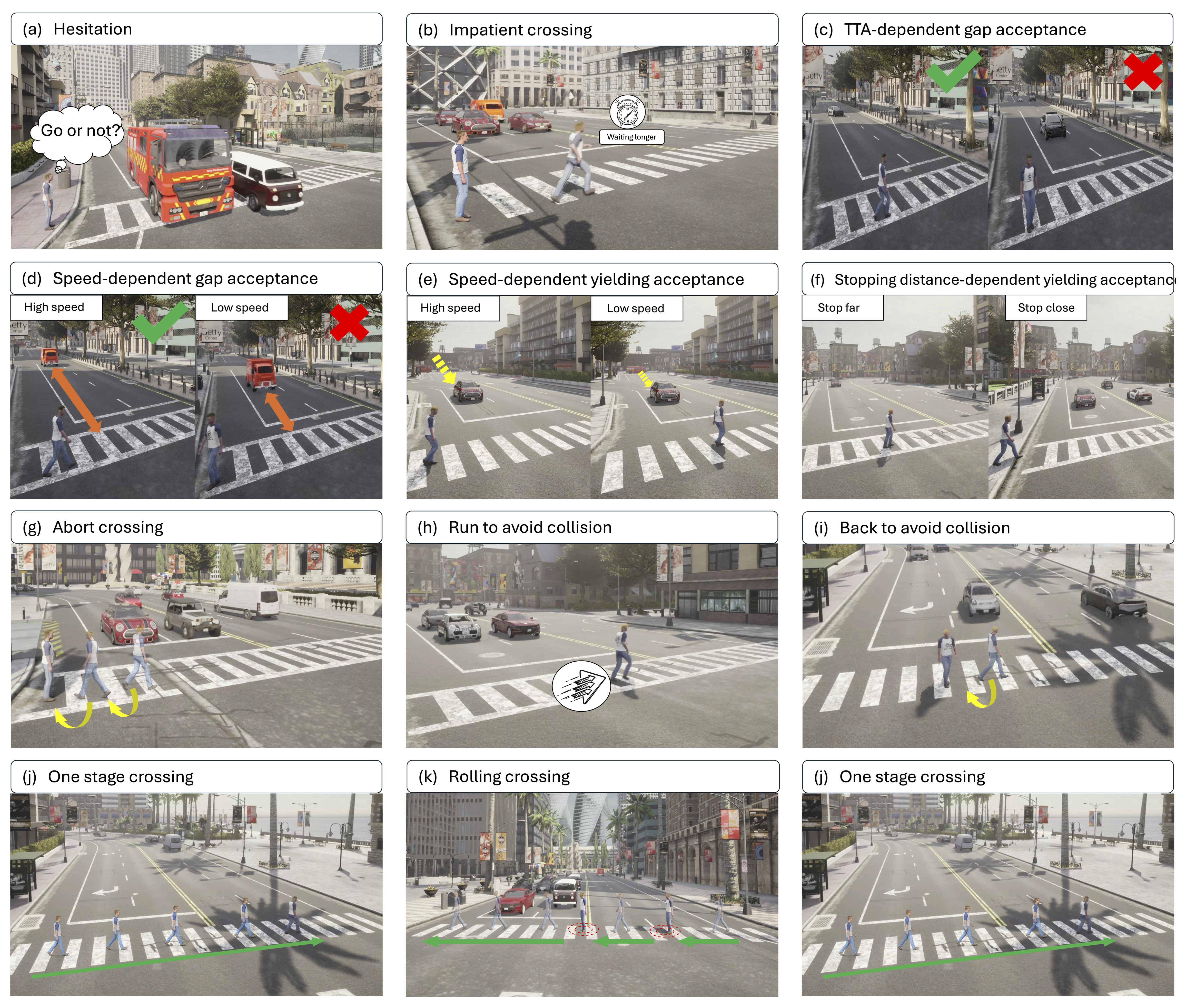}
    \caption{Human-like behaviors exhibited by the proposed pedestrian model during crossing.
    (a) Before crossing, the model hesitates and appraises approaching vehicles~\cite{gorrini2016towards}.
    (b) Longer waiting increases impatience and leads to the acceptance of smaller gaps~\cite{theofilatos2021cross}.
    (c)--(d) Gap acceptance increases with time to arrival (TTA) and, for the same TTA, is higher for faster approaching vehicles~\cite{petzoldt2014relationship,tian2022explaining}.
    (e)--(f) Under yielding conditions, a lower approach speed and a farther stopping position promote earlier crossing initiation~\cite{tian2023deceleration}.
    (g) The model may initiate a tentative crossing and then abort if danger is detected~\cite{balwan2017impacts}.
    (h)--(i) During crossing, the model may run or step back to avoid a collision~\cite{zhuang2011pedestrians}.
    (j)--(k) The model tends to use one-stage crossing in sparse traffic and rolling crossing in denser traffic~\cite{zhang2019evaluation}.
    (l) The model adapts to different traffic norms: in more yielding traffic, it exhibits more direct and less evasive behavior, while fine-tuning further improves its alignment with local traffic norms.}
    \label{fig1:human_like_behaviors}
\end{figure*}

We here propose a pedestrian model that reproduces human behavior and can be deployed in high-fidelity traffic simulations (Fig.~\ref{fig1:human_like_behaviors}). 

Our model builds on the theory of computational rationality (CR)~\cite{oulasvirta2022computational,gershman2015computational,howes2023towards}. Previous work has shown that this framework can model pedestrian crossing as a resource-bounded decision-making problem under perceptual uncertainty and cognitive constraints~\cite{wang2025modeling,wang2025pedestrian}.
It assumes that pedestrian behavior adapts according to their perceptual bounds, which determine the accuracy of their beliefs about their environments, their motor bounds, which define their ability to take action, and their rewards, which define what they regard as important.

In this work, we formulate pedestrian crossing as a POMDP~\cite{oulasvirta2022computational,wang2025pedestrian}, specifying the observations, actions, rewards, and belief estimates of a pedestrian based on empirical and modeling work on pedestrian behavior. 

The model receives bounded, task-relevant information about the pedestrian, crosswalk, destination, and nearby vehicles, and uses belief estimates of vehicle motion rather than perfect global knowledge to reflect perceptual uncertainty. 
Its action space includes stopping, walking, running, and stepping back~\cite{cynecki1980development,kadali2020evaluation}, allowing the pedestrian to update behavior throughout the crossing episode. 
Compared with the pedestrian model of Wang et al.~\cite{wang2025pedestrian}, which is limited to single lane, single vehicle, constant vehicle speed, and binary crossing response, our formulation supports multi-lane environments, multiple dynamically moving vehicles, and closed-loop interaction in continuously evolving traffic scenes.
Moreover, and in contrast to prior pedestrian simulations that rely on fixed gap thresholds~\cite{tian2022explaining,pekkanen2022variable,wang2025pedestrian} or pre-specified crossing rules~\cite{montali2023waymo,dosovitskiy2017carla}, our model learns from a reward function that encodes competing objectives balancing safety, crossing efficiency, waiting cost, and motor effort. 
This unified objective allows human-like crossing behaviors to emerge without explicit handcrafting, including changes in initiation, movement adjustments, retreats, and crossing strategies as traffic conditions vary.

After training with deep RL, the model can be deployed directly in a high-fidelity urban traffic simulator (i.e., CARLA) to develop and test automated vehicles. Given a start point and destination, it navigates toward the goal while producing human-like crossing behavior in interaction with surrounding vehicles.

The model reproduces three types of empirically documented crossing behaviors (Fig.~\ref{fig1:human_like_behaviors}): crossing initiation, adaptive behavior during crossing, and higher-level strategy selection. 
Rather than assessing these behaviors as fixed outputs, we examine whether their occurrence varies systematically with traffic density and vehicle approach dynamics, as reported in empirical studies of pedestrian crossings~\cite{theofilatos2021cross,wang2025pedestrian,almodfer2016quantitative,zhang2019evaluation}. We further evaluate safety outcomes against CARLA's AI walker, and test whether the learned policy transfers to unseen maps and adapts through lightweight fine-tuning to different traffic environments and local traffic norms.

\section{Related Work}
We organize this review around a behavior-based comparison of pedestrian models. We first identify ten representative crossing phenomena from the empirical literature and describe them across three stages of the crossing process: crossing initiation, adaptation during crossing, and higher-level crossing strategy selection. We then review representative existing pedestrian models and examine which of these phenomena they address. Table~\ref{tab:model_behavior_matrix} summarizes their behavioral coverage, generalization across traffic contexts, and suitability for closed-loop simulation.
\begin{table*}[t]
\centering
\caption{Coverage of behaviors, generalizability, and simulator usability across different pedestrian-crossing models.}
\label{tab:model_behavior_matrix}

\setlength{\tabcolsep}{4pt}
\renewcommand{\arraystretch}{1.12}

% \begin{tabular*}{\textwidth}{@{\extracolsep{\fill}}lcccccccccccc@{}}
\begin{tabular*}{\textwidth}{@{\extracolsep{\fill}}>{\centering\arraybackslash}p{0.26\textwidth}cccccccccccc@{}}
\toprule
\multirow[c]{2}{*}{\kern0mm Model} &
\multicolumn{6}{c}{Decisions to initiate crossing} &
\multicolumn{2}{c}{During crossing} &
\multicolumn{2}{c}{Strategies} &
\multirow{2}{*}{Gen.} &
\multirow{2}{*}{Sim.} \\
\cmidrule(lr){2-7}\cmidrule(lr){8-9}\cmidrule(lr){10-11}
& H & I & TTA & SpdGA & SpdYA & SDYA & Abort & RunBack & OneStage & Rolling &  &  \\
\midrule
Binary logit model~\cite{tian2022explaining}
&  & \cmark & \cmark & \cmark &  &  &  &  &  &  & \xmark & \xmark \\
Mechanistic models~\cite{pekkanen2022variable}
&  &  & \cmark & \cmark &  & \cmark &  &  &  &  & \xmark & \xmark \\
Strategy statistical models~\cite{opiew2023study}
&  &  &  &  &  &  &  &  & \cmark & \cmark & \xmark & \xmark \\
Bounded optimality model~\cite{wang2025pedestrian}
&  &  & \cmark & \cmark & \cmark & \cmark &  &  &  &  & \xmark & \xmark \\
Deep learning methods~\cite{fang2018pedestrian,wang2024multi}
&  &  & \cmark &  &  &  &  &  &  &  & \cmark & \xmark \\
Hierarchical forecasting model~\cite{yang2024hierarchical}
& \cmark &  &  &  &  &  &  & \cmark &  &  & \xmark & \xmark \\
Suicidal Pedestrian~\cite{yang2023suicidal}
&  &  &  &  &  &  &  &  &  &  & \xmark & \cmark \\
CARLA AI walker
& \cmark &  &  &  &  &  & \cmark &  & \cmark &  & \cmark & \cmark \\
\midrule
Ours
& \cmark & \cmark & \cmark & \cmark & \cmark & \cmark & \cmark & \cmark & \cmark & \cmark & \cmark & \cmark \\
\bottomrule
\end{tabular*}

\vspace{1mm}
\footnotesize
\textit{Behavior columns:}
H = Hesitation; I = Impatient crossing; TTA = TTA-dependent gap acceptance; SpdGA = Speed-dependent gap acceptance;
SpdYA = Speed-dependent yielding acceptance; SDYA = Stopping-distance--dependent yielding acceptance;
Abort = Abort crossing; RunBack = Run-and-back movement; OneStage = One-stage crossing; Rolling = Rolling crossing.
\textit{Gen.} = Generalizability; \textit{Sim.} = Simulator usability.
\end{table*}

%Empirical Crossing Phenomena
\subsection{Empirical Crossing Phenomena}
\label{sec:empirical_phenomena}

Empirical studies have documented a range of pedestrian crossing phenomena under different traffic conditions. We focus on ten representative phenomena and organize them into three stages of the crossing process: crossing initiation, adaptation during crossing, and higher-level crossing strategy selection.

\subsubsection{Decisions to initiate crossing}
The pedestrian’s decision to initiate crossing concerns whether and when to step into the roadway after assessing approaching vehicles. Empirical studies suggest that this stage is characterized by several behaviors. First, pedestrians often enter an appraising phase when approaching a crosswalk: they slow down, briefly pause, or stop while evaluating the distance and speed of oncoming vehicles before committing to cross, a process commonly interpreted as hesitation~\cite{cynecki1980development,gorrini2016towards}. Second, longer waiting times tend to increase pedestrians’ willingness to cross, leading them to accept shorter gaps and exhibit riskier crossing behavior, a phenomenon often described as impatience~\cite{chen2017impact,theofilatos2021cross,arman2015pedestrian}. 

A large body of work has also examined gap-acceptance behavior at crossing initiation. The probability of accepting a gap generally increases with the TTA of an approaching vehicle~\cite{petzoldt2014relationship}. For a given initial TTA, acceptance decisions are further shaped by vehicle speed, with some studies reporting higher acceptance rates for faster vehicles~\cite{lobjois2007age,tian2022explaining}. In yielding scenarios, pedestrians often interpret lower approach speed as an implicit cue of yielding, especially when the vehicle is closer to the crosswalk~\cite{tian2023deceleration}. Similarly, when a vehicle stops farther from the crosswalk, pedestrians are more likely to perceive stronger yielding intent and initiate crossing~\cite{tian2023deceleration}.

\subsubsection{Decisions during crossing}
Decisions during crossing concern how pedestrians respond after they have already entered the roadway, especially when the situation becomes unsafe. Empirical studies report two common behaviors at this stage. First, pedestrians may abort a tentative crossing and retreat to the curb if they judge the available gap to be too small or risky~\cite{balwan2017impacts}. Second, when danger arises during crossing, they may either run forward or step back to move away from the potential collision region~\cite{zhuang2011pedestrians}.

\subsubsection{Crossing strategies}
Crossing strategies describe how pedestrians cross a multi-lane roadway, either by crossing all lanes in one movement or by proceeding lane by lane with possible waiting in intermediate safe positions. Empirical studies suggest that pedestrians tend to adopt one-stage crossing under light traffic, but switch to rolling crossing under denser or more complex conditions when no sufficiently large overall gap is available~\cite{zhang2019evaluation}.

%Pedestrian Crossing Models
\subsection{Pedestrian Crossing Models}

Pedestrian crossing models differ in their primary objectives and behavioral coverage. We group representative models into explanatory, predictive, and simulation-oriented approaches, and compare which of the ten crossing phenomena described above they address.

\subsubsection{Explanatory models}

Explanatory models aim to characterize the factors and mechanisms underlying pedestrian crossing decisions. Statistical and discrete-choice models estimate the probability of crossing initiation from variables such as waiting time, vehicle time to arrival, vehicle speed, and pedestrian characteristics~\cite{petzoldt2014relationship,tian2022explaining,arman2019applied}. These models can account for empirical relationships such as increased willingness to cross after prolonged waiting and TTA- or speed-dependent gap acceptance. Statistical models have also been used to explain the selection of higher-level strategies, such as one-stage versus rolling crossing, based on traffic conditions and pedestrian attributes~\cite{opiew2023study}. Although interpretable, these models typically focus on individual discrete decisions rather than continuous behavioral adjustment throughout the crossing process.

Mechanistic and cognitively grounded models represent the processes through which crossing decisions arise. Evidence-accumulation models, including variable-drift diffusion models, describe how noisy perceptual evidence develops over time until a crossing decision is reached~\cite{pekkanen2022variable,ratcliff2016diffusion}. Bounded-optimality models grounded in computational rationality explain pedestrian behavior as adaptation under perceptual, cognitive, and motor constraints~\cite{wang2025pedestrian}. Game-theoretic models provide another explanatory perspective by representing pedestrian and vehicle decisions as interdependent strategic choices~\cite{amini2021towards}. These models provide explicit accounts of how uncertainty and human constraints shape crossing behavior. However, existing implementations have primarily been evaluated for crossing initiation or in simplified interaction settings, rather than for continuous behavior throughout complex, multi-lane crossing scenarios.

\subsubsection{Predictive models}

Predictive models aim to infer pedestrian crossing intention or forecast future motion from observational data. Deep learning methods learn these relationships from image, video, or multimodal datasets of pedestrian--vehicle interaction~\cite{rasouli2017they,fang2018pedestrian}. Convolutional neural networks, recurrent neural networks, and Transformer-based architectures predict crossing intention or short-horizon trajectories from pedestrian appearance, pose, motion history, and surrounding scene context~\cite{razali2021pedestrian,yang2022predicting,wang2024multi,yin2021multimodal}. Hierarchical forecasting models additionally represent pedestrian behavior at multiple levels, combining high-level behavioral states, such as waiting, walking, and running, with low-level motion prediction~\cite{yang2024hierarchical}. More recent vision-language and LLM-based approaches explore the use of semantic scene understanding and contextual reasoning for pedestrian--vehicle interaction~\cite{pu2026vision}.

These models can capture complex relationships in observational data and provide predictions in recorded traffic scenes. However, they are primarily evaluated in terms of intention-classification or trajectory-prediction accuracy, rather than their ability to reproduce empirically documented behaviors. Their performance can also decrease under shifts between datasets, locations, or traffic contexts~\cite{zhang2023pedestrian,gesnouin2022assessing}.

\subsubsection{Simulation-oriented pedestrian models}

Beyond the computational models reviewed above, simulation-oriented pedestrian models are widely used in automated-driving pipelines for training, testing, and evaluation. Representative directions include safety-critical pedestrian agents for stress testing~\cite{priisalu2022generating,yang2023suicidal}, simulator-native controllers such as CARLA’s AI walker and SVL’s pedestrian agents~\cite{lyssenko2021evaluation,lu2025can}, controllable motion-generation systems for interactive testing~\cite{rempe2023trace}, and industry simulation agents for AV training and evaluation~\cite{montali2023waymo}.

Although useful for scalable simulation and benchmarking, these models are generally not designed to systematically reproduce empirically documented pedestrian crossing mechanisms under bounded human constraints. Their human-likeness, behavioral coverage, and generalization across interaction contexts therefore remain limited. As summarized in Table~\ref{tab:model_behavior_matrix}, existing approaches typically capture only a subset of the behaviors considered in this paper or remain difficult to align with interpretable pedestrian decision mechanisms. To address this gap, we propose a theory-grounded, simulator-ready pedestrian model for closed-loop interaction in complex traffic.

\section{Methodology}

\subsection{Pedestrian Model}

\begin{figure*}[!t]
    \centering
    \includegraphics[width=\textwidth]{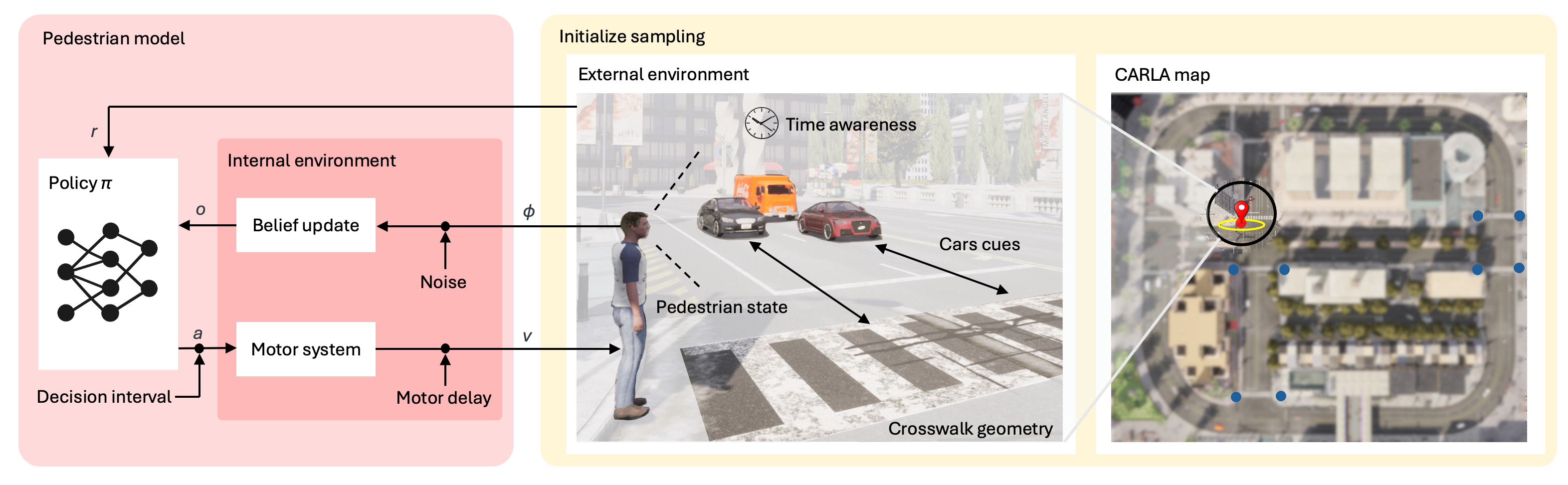}
    \caption{Architecture of the proposed closed-loop pedestrian model. At the beginning of each training episode, a crosswalk is sampled from the map, and the pedestrian is initialized at one side with the opposite side specified as the destination. The pedestrian receives bounded and noisy perceptual input \(\phi\) about nearby vehicles, together with task-related information about the pedestrian and crosswalk. A constant-acceleration Kalman filter (CAKF) estimates vehicle motion from the noisy measurements, and the resulting estimates are combined with the task information to form the policy input \(o\). At each decision update, the policy \(\pi\) maps \(o\) to a discrete locomotion action \(a\). The motor system applies execution delay and speed constraints to produce a velocity command \(v\). CARLA executes the command and advances the traffic state, completing the interaction loop. During training, the reward \(r\) is used to update the policy parameters.}
    \label{fig:model_architecture}
\end{figure*}

The proposed model generates pedestrian crossing behavior through closed-loop interaction with the CARLA traffic simulator~\cite{dosovitskiy2017carla}. Building on computational rationality~\cite{oulasvirta2022computational,wang2025pedestrian}, pedestrian crossing process is formulated as sequential decision making under perceptual, cognitive, and motor constraints. Rather than making a single decision before roadway entry, the model repeatedly perceives surrounding traffic, estimates vehicle motion, selects a locomotion action, and revises its behavior as the traffic situation evolves. Fig.~\ref{fig:model_architecture} shows the architecture of pedestrian model.

During crossing, CARLA maintains the complete traffic state, including the pedestrian, crosswalk geometry, and surrounding vehicles. However, the pedestrian does not have direct access to this state and receives only partial and noisy information about the environment. After belief updating, the resulting representation \(o\) is provided to the policy, which selects one of four locomotion actions: \texttt{stop}, \texttt{walk}, \texttt{run}, or \texttt{back}. This closed-loop process allows the pedestrian to revise its actions as the surrounding traffic evolves.  Representative closed-loop interactions generated by the model are shown in the supplementary video.

Human constraints are incorporated at different stages of the architecture. Perceptual constraints limit the amount and accuracy of vehicle information available to the policy. A fixed decision-update interval represents the intermittent nature of human action selection, while motor delay and speed limits constrain action execution. During training, the policy is optimized through reinforcement learning using rewards collected from repeated interactions with the simulator. The following subsections describe the human constraints and POMDP formulation.

\subsubsection{Perceptual, cognitive, and motor constraints}

\paragraph{Noisy perception}
We made two assumptions about pedestrian perception. First, the pedestrian could not observe all vehicles in the scene and instead perceived only a limited subset of surrounding traffic. Second, following Markkula et al.~\cite{markkula2023explaining}, we assumed that visual distance estimation was corrupted by Gaussian angular noise with standard deviation $\sigma_v$.

We set pedestrian eye height to $1.6\,\mathrm{m}$~\cite{sudkamp2021role} and the angular-noise standard deviation to $\sigma_v = 0.1$. Because humans are less sensitive to acceleration~\cite{brouwer2002perception}, we used a CAKF as a belief-update module to estimate vehicle position, velocity, and acceleration from noisy distance observations. These estimated kinematic states were then provided to the policy as part of the observation input.

\paragraph{Decision update interval}
Human action selection is not updated continuously at arbitrarily fine time scales, but is often better described as intermittent adjustment~\cite{gawthrop2011intermittent,markkula2018sustained}. To capture this limited temporal resolution, the policy updated its action every $\Delta t_d = 0.3\,\mathrm{s}$. At each decision time, the pedestrian updated its belief using the latest perceptual observation and selected an action based on the updated belief. This value was chosen based on prior reports of inter-adjustment intervals on the order of a few hundred milliseconds~\cite{loram2012identification}.

\paragraph{Motor constraints}
After an internal decision is made, motor commands require finite time to propagate from the brain to the limbs and produce overt movement. We modeled two motor constraints: motor delay and physical speed limits.

\emph{Motor delay}.
Experimental studies have shown that lower-limb motion typically begins several hundred milliseconds after stimulus onset~\cite{mackinnon2007preparation,jaworski2013estimation}. To capture this constraint, whenever the pedestrian changed locomotion state, the new action was not executed immediately but only after a stochastic delay. Based on prior measurements of motor latency~\cite{darbutas2013dependence}, we modeled this delay as Gaussian with mean \(0.35\,\mathrm{s}\) and standard deviation \(0.15\,\mathrm{s}\).

\emph{Physical speed limits}.
Pedestrian locomotion was constrained by plausible human movement speeds. We therefore assigned action-dependent target speeds of \(0\,\mathrm{m/s}\) for \texttt{stop}, \(1.2\,\mathrm{m/s}\) for \texttt{walk}~\cite{mutcd2003_part4e,alam2025pedestrian}, \(3.0\,\mathrm{m/s}\) for \texttt{run}~\cite{selinger2022running}, and \(-2.0\,\mathrm{m/s}\) for \texttt{back}. The running speed corresponds to a moderate pace, whereas backward motion was capped below running speed to reflect its role as an evasive maneuver.

\subsubsection{POMDP formulation}
Given the human constraints specified above, we formulate pedestrian crossing as a POMDP and learn a bounded-optimal policy under partial observability.
We use the compact POMDP specification $\langle S, A, T, O, R \rangle$, where $S$ is the state space, $A$ is the action space, $T$ is the transition function, $O$ is the observation space, and $R$ is the reward function.

\paragraph{State}
In our setting, the environment at each time step $t$ is in a latent state $s_t \in S$, which includes the pedestrian’s current position and velocity, as well as the positions, velocities, and accelerations of all vehicles in the scene. In addition, the states also include the attributes of crosswalks, such as the number of lanes and its length.  

\paragraph{Observation}
The observation space ${O}$ is designed to approximate what a human pedestrian could realistically perceive. We restrict the observation to the two closest approaching vehicles per lane within \(70\,\mathrm{m}\) of the pedestrian~\cite{petzoldt2014relationship}, reflecting pedestrians’ bounded perceptual and attentional resources. The leading vehicle typically dominates the immediate gap-acceptance constraint~\cite{shaaban2020pedestrian}, while the second vehicle captures near-future threats after the leader passes~\cite{theofilatos2021cross}. Beyond the second vehicle, visibility is often reduced by occlusion and the marginal utility of additional vehicles diminishes~\cite{zhu2021interactions}, making this a reasonable bounded-optimal approximation of human perception. Once the leading vehicle in a lane has passed the crosswalk, it is removed from the observation, and the next vehicle in that lane (the former third vehicle) becomes observable.

For each observed vehicle, the pedestrian receives:
\begin{itemize}
    \item noisy estimates of vehicles' longitudinal distance, velocity and acceleration, obtained from a CAKF applied to the noisy distance measurements.
\end{itemize}

Additionally, the observation also includes following task-related variables:
\begin{itemize}
    \item a time-awareness signal (elapsed time since the start of crossing);
    \item the total length of the crosswalk;
    \item the index of the lane currently occupied by the pedestrian;
    \item the distance to the next lane boundary;
    \item the distance to the goal (end point of the crosswalk);
    \item the total number of lanes of the current crosswalk;
    \item the pedestrian's current walking speed.
\end{itemize}

\paragraph{Action Space}
The action space $\mathcal{A}$ is discrete and contains four locomotion commands:
\begin{equation}
    \mathcal{A} = \{\texttt{stop}, \texttt{walk}, \texttt{run}, \texttt{back}\}.
\end{equation}
These actions correspond to the target speeds specified in the previous subsection (standing, walking, running, and moving backwards). The pedestrian’s walking direction follows the crosswalk direction, i.e., from the crosswalk entry to the exit.

\paragraph{Transition}
Transitions were implemented by advancing CARLA in synchronous mode with a fixed simulation step of \(\Delta t_s=0.1\,\mathrm{s}\). Vehicle states followed CARLA Traffic Manager dynamics, crosswalk geometry remained fixed within each episode, and pedestrian motion was updated from the executed locomotion command subject to speed limits and delayed actuation. The delayed motor execution state was included in the latent state to preserve the Markov property.

\paragraph{Reward}
Pedestrian crossing can be viewed as a trade-off between safety and efficiency: pedestrians aim to complete the crossing as quickly as possible while avoiding collisions and overly risky interactions \cite{tian2022explaining}. Consistent with this view, we organize the reward function $R$ into three parts: core task rewards, behavioral preference rewards, and shaping rewards.

\emph{Core task rewards}.
The first part of the reward encodes the basic objective of crossing safely and efficiently.
\begin{itemize}
    \item {Success reward:} when the pedestrian successfully reaches the opposite sidewalk, a terminal reward of \(+40\) is given.
    \item {Collision penalty:} if the pedestrian is hit by a vehicle at any time, a terminal penalty of \(-120\) is applied. Together with the success reward, this establishes a safety-first task objective.
    \item {Time cost:} at every simulation step of \(\Delta t_s = 0.1\,\text{s}\), the agent receives a time penalty of \(-0.02\), which encourages timely crossing.
\end{itemize}

\emph{Behavioral preference rewards}.
The second part of the reward is grounded in basic principles of human behavior and is designed to improve behavioral realism. 
\begin{itemize}
    \item {Near-miss penalty:} pedestrians are sensitive to approaching or dangerously close vehicles and tend to avoid such interactions ~\cite{delucia2008critical}. We introduce an additional penalty of \(-3\) whenever the distance between the pedestrian and any vehicle becomes smaller than \(3\,\text{m}\) without an actual collision. This reward provides a coarse reward-level proxy for aversion to threatening proximity~\cite{tian2022explaining}.
    \item {Action cost:} to discourage unnecessarily effortful maneuvers and improve behavioral realism, we add an action-dependent locomotion cost motivated by evidence that humans prefer walking speeds close to the minimum cost of transport~\cite{ralston1958energy,majed2024walking}. Accordingly, \textit{stop} and \textit{walk} incur no extra cost, whereas \textit{run} incurs \(-0.5\) and \textit{back} incurs \(-0.1\). This biases the agent toward walking when safe, while still allowing running or retreating when needed.
\end{itemize}

\emph{Reward shaping}.
Finally, to reduce reward sparsity and stabilize optimization, we include a progress-based shaping reward:
\begin{itemize}
    \item {Progress reward:} let \(p \in [0,1]\) denote the fraction of the crosswalk that has been completed. For each additional \(1\%\) of progress, the agent receives a reward of \(+0.05\), so that fully traversing the crosswalk yields a total progress reward of \(+5\).
\end{itemize}

\subsection{Training and Simulation Setup}

\subsubsection{CARLA environment}

We train the pedestrian policy in the CARLA driving simulator~\cite{dosovitskiy2017carla}. Training is conducted in \textit{Town10HD}, which contains marked crosswalks with different road geometries, including two- and four-lane roads, four-way intersections, and T-junctions. CARLA is operated in synchronous mode with a fixed simulation step of \(\Delta t_s=0.1\,\mathrm{s}\), ensuring consistent timing between vehicle motion, pedestrian control, and policy interaction.

Environmental conditions, including weather, illumination, and traffic-light timing, are held fixed during training to control sources of variation outside the scope of this study. Background vehicles are controlled by the CARLA Traffic Manager and obey traffic signals. Pedestrian yielding is disabled so that the policy cannot rely on approaching vehicles consistently giving way. Vehicles may nevertheless decelerate or stop because of traffic signals, congestion, or interactions with other vehicles, resulting in a range of approach and braking patterns near the crosswalks.

The pedestrian policy does not receive the traffic-light phase as part of its observation. This design isolates crossing decisions driven by perceived vehicle motion and interaction risk rather than explicit signal compliance. Modeling pedestrian responses to traffic signals and their interaction with other social or contextual factors is outside the scope of the present work.

\subsubsection{Domain randomization}

We apply domain randomization to expose the policy to diverse road geometries and pedestrian--vehicle encounters during training. At the beginning of each episode, a marked crosswalk is randomly selected from \textit{Town10HD}. The pedestrian is initialized at one side of the crosswalk, and the opposite side is specified as the destination. Because the sampled crosswalks differ in lane count, length, location, and intersection layout, the policy experiences multiple crossing configurations within the same training map.

The map is populated with \(35\)--\(45\) vehicles sampled from the CARLA vehicle blueprint library. Vehicle target speeds are sampled between \(20\) and \(50\,\mathrm{km/h}\), while the evolving traffic flow produces variation in vehicle arrival times, local traffic density, and interactions among vehicles. The combination of crosswalk sampling, vehicle-type variation, randomized vehicle speeds, and evolving traffic conditions reduces dependence on particular vehicle trajectories or crossing locations.

Each episode lasts for at most 600 simulation steps, corresponding to \(60\,\mathrm{s}\) of simulated time. An episode terminates when the pedestrian reaches the destination, collides with a vehicle, or exceeds the time limit. These outcomes are respectively recorded as successful crossing, collision, and no crossing.

\subsubsection{Reinforcement learning}

Given the POMDP formulation above, we use model-free deep RL learning to learn a bounded-optimal crossing policy. Concretely, the crossing scenario is wrapped as a Gymnasium-compatible environment, which interfaces the simulator with Python and exposes the observations, actions, and rewards defined in the POMDP. The pedestrian policy is parameterized as \(\pi_\theta(a_t \mid o_t)\) and trained to maximize the expected discounted return.

We adopt Proximal Policy Optimization (PPO)~\cite{schulman2017proximal} as the optimization algorithm. The main training parameters are a learning rate of \(3\times10^{-4}\), batch size of \(64\), discount factor \(\gamma = 0.99\), clipping range of \(0.2\), and entropy coefficient of \(0\). 

Once trained, the model can be deployed in simulation by specifying a start and a goal location, and can be further adapted to different maps and traffic conditions through continued training to obtain pedestrian models for specific environments.

\subsection{Evaluation Protocol}
\label{sec:evaluation_protocol}

The evaluation addresses three questions: whether the model reproduces empirically documented crossing behaviors, whether it operates effectively as a closed-loop pedestrian agent, and whether its behavior transfers and adapts across traffic environments. All experiments were conducted in CARLA under closed-loop pedestrian--vehicle interaction. Unless otherwise specified, the trained policy was evaluated without additional learning. Each episode initialized the pedestrian at one side of a crosswalk and specified the opposite side as the destination.

For dynamic traffic analyses, traffic density was defined as the number of unique vehicles passing within a \(15\,\mathrm{m}\) radius of the crosswalk during an episode. Episodes were categorized as low density when no more than two nearby vehicles were observed, medium density when three to seven vehicles were observed and high density when at least eight vehicles were observed. The same density definition was used for the behavioral and closed-loop outcome evaluations.

\subsubsection{Behavioral evaluation}

To assess behavioral realism, we evaluate whether the trained policy reproduces ten representative pedestrian crossing behaviors reported in prior literature. These behaviors cover three stages of decision making: crossing initiation, adaptation during crossing, and higher-level crossing strategy selection. Most behaviors are evaluated in dynamic traffic under predefined traffic-density conditions, with randomized crosswalk initialization and continuously moving Traffic Manager-controlled vehicles. Controlled scenarios are used to isolate the effects of waiting time, vehicle TTA, vehicle speed, and yielding behavior. To distinguish the decision stages, we define the threshold line as the boundary between the sidewalk and roadway along the crosswalk. Behavior occurring before the pedestrian crosses this line is treated as crossing-initiation behavior, whereas behavior occurring after roadway entry is treated as behavior during crossing. Where comparable quantitative human data are available, we compare the model results with reported human distributions or trends. Otherwise, behavioral validity is assessed by whether the model reproduces the direction of the qualitative effect documented in prior studies.

Crossing initiation behaviors included hesitation, impatient crossing, TTA-dependent gap acceptance, speed-dependent gap acceptance, speed-dependent yielding acceptance and stopping distance-dependent yielding acceptance. Hesitation was defined as stopping or briefly moving backward before entering the roadway, and was analyzed as a function of traffic density. Impatient crossing was tested using repeated identical gaps and quantified as the cumulative probability of first gap acceptance over waiting time. Gap acceptance was measured as roadway entry before an approaching vehicle reached the crosswalk, with acceptance probabilities compared across TTA and vehicle-speed conditions. Yielding acceptance was evaluated in controlled braking scenarios by measuring crossing initiation time when vehicles stopped before the crosswalk, with comparisons across approach speed and stopping distance.

Behaviors during crossing were evaluated after the pedestrian had crossed the threshold line and entered the roadway. Abort crossing was defined as entering the roadway and subsequently retreating to the safe region within the same crossing attempt. Run/back evasive movement was quantified by the use of \emph{run}, \emph{back} and stopping related adjustments during crossing, especially under dense traffic and dangerous situations with an approaching vehicle TTA below \(5\,\mathrm{s}\). 

Higher level crossing strategy was evaluated by distinguishing one-stage crossing, in which the pedestrian completed the crossing using only forward movement actions, from rolling crossing, in which the pedestrian interrupted or segmented the crossing using actions such as \emph{stop} or \emph{back}. Human-like patterns were defined according to empirical trends reported in prior work, such as higher hesitation, abort and rolling-crossing rates under denser traffic, higher gap acceptance for longer TTA, and earlier yielding acceptance for slower or earlier-stopping vehicles.

\subsubsection{Closed-loop outcome evaluation}

To assess simulator usability and safety-related performance, we evaluated the trained model in dynamic traffic using three mutually exclusive episode level outcomes: successful crossing, no crossing and collision. Successful crossing was defined as reaching the opposite side of the crosswalk before the time limit; no crossing as reaching the time limit without success or collision; and collision as any contact with a vehicle during the episode. Outcome rates were computed overall and stratified by traffic density. CARLA's AI walker baseline was evaluated under the same dynamic traffic setting using the same outcome definitions.

\subsubsection{Transfer and adaptation evaluation}

To assess generalization, we evaluated the pretrained policy on an unseen CARLA map under broadly similar traffic conditions, first without additional training and then after continued training in the target map. To assess adaptation to different traffic norms, we further evaluated the model in a modified traffic environment with lower vehicle speed, lower traffic density and stronger vehicle yielding behavior. For both evaluations, we compared original-domain performance, zero-shot transfer and fine-tuned target-domain performance using the same outcome categories. We also analyzed changes in behavioral composition, including hesitation, abort crossing, rolling crossing and action proportions, to determine whether adaptation affected both safety outcomes and qualitative crossing style.

\subsection{Statistical Analysis}

Unless otherwise noted, the primary unit of analysis was the episode. This choice was appropriate because most behavioral annotations and outcome measures were defined at the episode level, such as whether a trial contained hesitation, abort crossing, rolling crossing, no crossing, or collision. For these binary outcomes, we used logistic regression with traffic density coded as an ordered predictor (Low = 1, Medium = 2, High = 3). This model was used because the outcomes were binary and because our main question was whether each behavior showed a monotonic trend across increasing traffic density, rather than whether all density levels differed from one another independently. We report odds ratios (ORs), 95\% confidence intervals (CIs), and two sided \(P\) values. For action-level analyses of \textit{run} and \textit{back} usage across traffic density conditions, we used the same approach, treating each action indicator as a binary event within the corresponding analysis unit.

For proportion estimates shown in figures, 95\% Wilson confidence intervals were used where applicable. Wilson intervals were chosen because they provide more stable uncertainty estimates for binomial proportions than the normal approximation, particularly when proportions are close to 0 or 1 or when group sizes differ across conditions. Descriptive analyses, such as cumulative first crossing curves in the impatient crossing test, were summarized directly from the empirical distributions because these analyses were intended to characterize the temporal pattern of acceptance rather than test a single binary contrast. Gap acceptance and yielding acceptance results were summarized as acceptance proportions or cumulative initial crossing curves across experimentally defined conditions and, when applicable, compared qualitatively with prior human data.

Sample sizes (\(N\)), reported \(P\) values, and the specific statistical test used for each experiment are provided in the corresponding figure legends or Results text. Statistical analyses were performed in Python using standard scientific computing libraries.

\section{Results}

Following the evaluation protocol described in Section~\ref{sec:evaluation_protocol}, we first examine whether the trained policy reproduces ten empirically documented pedestrian crossing behaviors. We then evaluate its closed-loop crossing outcomes and compare them with CARLA's AI walker, before examining zero-shot transfer and adaptation to new traffic environments.

\subsection{Behavioral Evaluation}

Across the ten target behaviors, the trained policy generally produced systematic changes in the same direction as the empirical regularities summarized in Section~\ref{sec:empirical_phenomena}. Behaviors that could be observed in randomized dynamic traffic without behavior-specific scenario construction—including hesitation, abort crossing, run and back actions, and one-stage versus rolling crossing—were evaluated over 6{,}000 episodes. Behaviors requiring the isolation of specific traffic factors were evaluated in controlled scenarios: impatient crossing was tested over 100 trials, while TTA- and speed-dependent gap acceptance and speed- and stopping-distance-dependent yielding acceptance were tested over 100 trials per condition. Crossing initiation varied with waiting time, vehicle time to arrival (TTA), vehicle speed, and yielding cues; after roadway entry, the policy used abort, running, and backward actions in response to changing traffic conditions; and its multi-lane crossing strategy varied with traffic density. Figs.~\ref{fig3:behavior_validation}--\ref{fig5:during_and_strategy} present the corresponding results, organized into crossing-initiation behaviors, behaviors during crossing, and crossing strategies.

\subsubsection{Crossing initiation behaviors}

\begin{figure}[!t]
    \centering
    \includegraphics[width=\columnwidth]{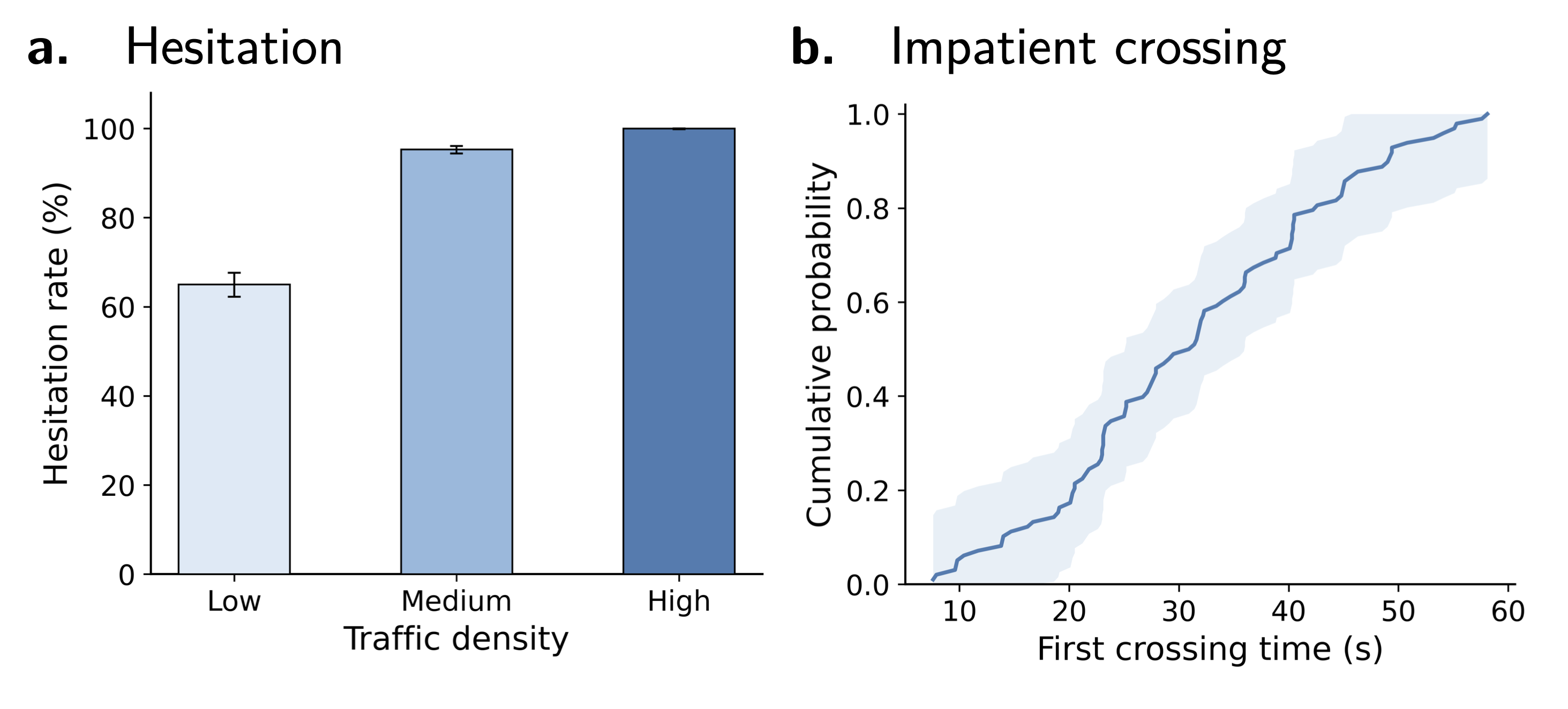}
    \caption{Model reproduces human-like initial crossing behaviors.
    (a) Proportion of episodes exhibiting hesitation across traffic-density conditions. Hesitation increases with traffic density, consistent with prior findings that pedestrians hesitate more under denser traffic with fewer available gaps~\cite{theofilatos2021cross}. Error bars denote 95\% confidence intervals.
    (b) Cumulative distribution function (CDF) of first crossing time in the impatient-crossing test. The increasing curve shows that the model progressively accepts the same gap after longer waiting, consistent with the impatience effect reported in pedestrian crossing studies~\cite{theofilatos2021cross}. Shaded bands denote 95\% confidence intervals.}
    \label{fig3:behavior_validation}
\end{figure}

\begin{figure}[!t]
    \centering
    \includegraphics[width=\columnwidth]{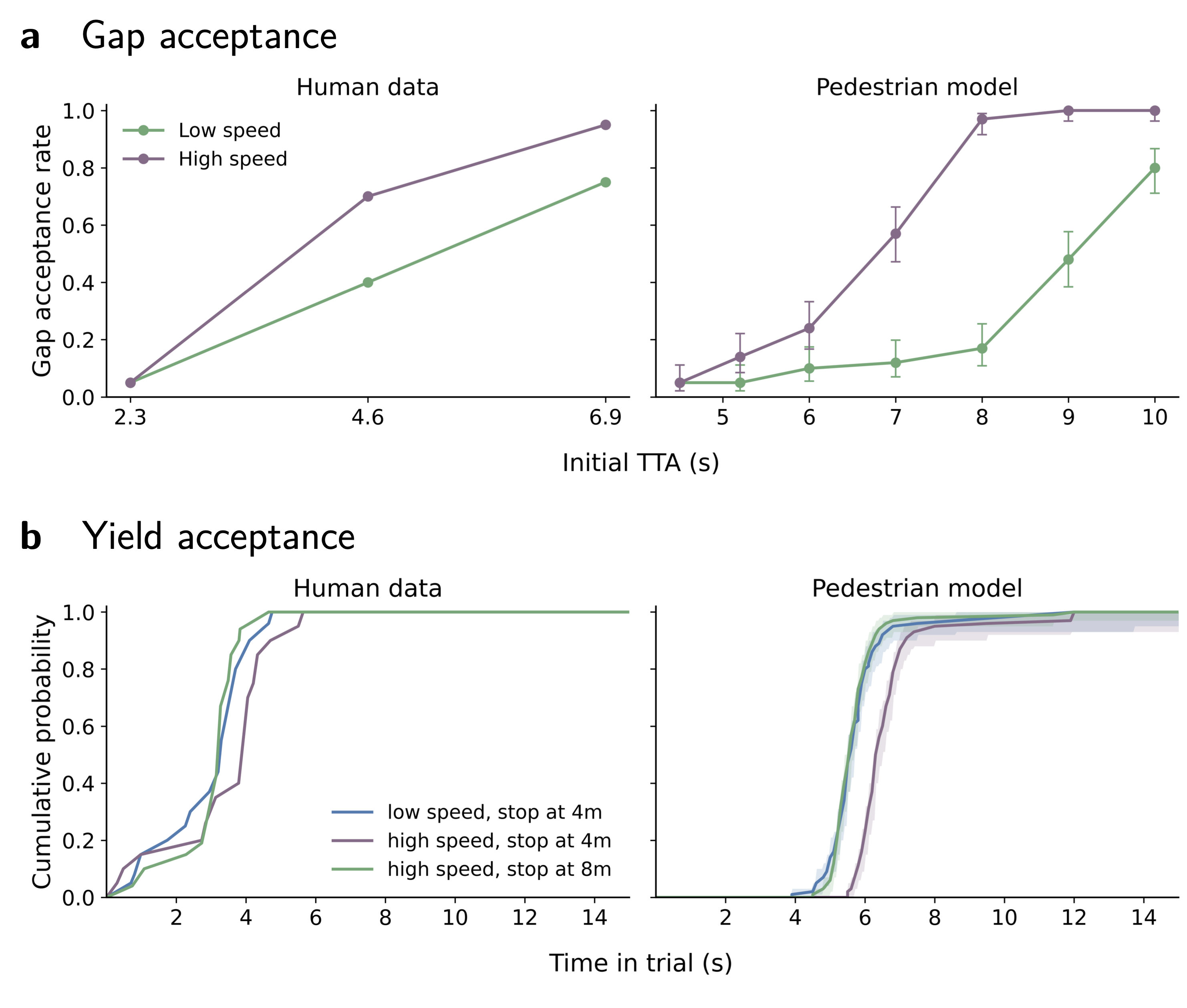}
    \caption{Results for gap acceptance test.
    (a) Gap acceptance under different initial TTAs and vehicle speed conditions, comparing human data~\cite{wang2025pedestrian} (left) and pedestrian model results (right). In both cases, acceptance increases with TTA and is generally higher at higher vehicle speed for a fixed TTA. Error bars denote 95\% Wilson binomial confidence intervals across 100 simulated trials per condition.
    (b) Yield-acceptance behavior under matched initial TTA, comparing human data~\cite{wang2025pedestrian} (left) and model results (right). Curves are shown for three conditions; earlier rising curves indicate earlier crossing initiation. For the same stopping distance (4 m), initiation occurs earlier in the low-speed condition; for the same vehicle speed (high speed), initiation occurs earlier when braking begins earlier and the stopping distance is larger. Shaded bands in the pedestrian model panel denote 95\% bootstrap confidence bands across 100 simulated trials per condition.}
    \label{fig4:TTA_behavior_validation}
\end{figure}

The results for the six crossing initiation behaviors are presented in Figs.~\ref{fig3:behavior_validation} and~\ref{fig4:TTA_behavior_validation}. These analyses examine how the policy approaches roadway entry and responds to waiting time, vehicle approach dynamics, and yielding cues.

\paragraph{Hesitation}
We label an episode as exhibiting hesitation if the pedestrian performs a pause (\textit{stop}) for at least one decision interval or a brief retreat (\textit{back}) before crossing. We evaluate this behavior under three traffic density conditions, with results summarized in Fig.~\ref{fig3:behavior_validation}(a). We observe that in most episodes the model typically exhibits a hesitation phase before crossing, and the hesitation rate increases with traffic density. Binary logistic regression with traffic density coded as an ordered predictor (Low=1, Medium=2, High=3) confirms a strong increasing trend (odds ratio per density level \(=11.7\), 95\% CI [10.1, 13.7], \(p<0.001\)). This pattern is consistent with prior empirical findings on density-dependent hesitation~\cite{theofilatos2021cross}.

\paragraph{Impatient crossing}
To test whether the pedestrian model exhibits impatient crossing, we place the simulated pedestrian at the curb and repeatedly present approaching vehicles with the same initial TTA, chosen such that the corresponding gap is initially rejected by the model. If it eventually accepts this gap after waiting for some time, this indicates that the agent becomes impatient and changed its gap acceptance behavior due to waiting time. As shown in Fig.~\ref{fig3:behavior_validation}(b), the cumulative probability of first acceptance increases over waiting time. Overall, the result suggests that the model captures the characteristic increase in impatience with waiting, in line with prior empirical findings~\cite{theofilatos2021cross}.

\paragraph{TTA-dependent and speed-dependent gap acceptance.}
TTA-dependent gap acceptance refers to the tendency that pedestrians are more likely to accept a gap when the TTA of an approaching vehicle is larger~\cite{petzoldt2014relationship}. Human data from Wang et al.~\cite{wang2025pedestrian} (Fig.~\ref{fig4:TTA_behavior_validation}(a), left) show an increase in gap acceptance with different initial TTAs. Using the same rationale (see \cite{wang2025pedestrian}), we varied vehicle speed and initial distance to construct multiple TTA conditions. As shown in Fig.~\ref{fig4:TTA_behavior_validation}(a), right, the model reproduces the same overall pattern: acceptance probability increases with TTA. Although the exact TTA ranges and crossing timing are not identical to those in Wang et al.~\cite{wang2025pedestrian}, partly because our setup uses different pedestrian starting positions and simulator-based vehicles with speed fluctuations during acceleration and normal driving rather than strictly controlled constant speed motion, the qualitative trend remains consistent with that observed in human data.

Even at the same TTA, pedestrians may exhibit different acceptance tendencies depending on vehicle speed~\cite{tian2022explaining}. As shown in Fig.~\ref{fig4:TTA_behavior_validation}(a), both the human data and the model results show higher acceptance rates under the higher-speed condition for comparable TTA levels. This indicates that the model captures the speed-dependent characteristic of pedestrian gap acceptance.

\paragraph{Speed-dependent and stopping distance-dependent yielding acceptance.}
To examine yielding-related crossing decisions, we use the same paradigm as Wang et al.~\cite{wang2025pedestrian}, comparing three vehicle conditions under same initial TTA: low speed, stop at 4 m; high speed, stop at 4 m; and high speed, stop at 8 m. The human data shown in Fig.~\ref{fig4:TTA_behavior_validation}(b), left are taken from Wang et al.~\cite{wang2025pedestrian}, who collected pedestrians' crossing-initiation times in a controlled experiment under these conditions. Under the same stopping distance (4 m), the blue curve rises earlier than the purple curve, indicating that pedestrians initiate crossing earlier when the vehicle approaches at lower speed and from a shorter distance. As shown in Fig.~\ref{fig4:TTA_behavior_validation}(b), right, the pedestrian model reproduces the same qualitative pattern. As noted above, the exact crossing-initiation times are not identical because of differences in pedestrian starting position and the simulator-based vehicle configuration, but the overall trend remains consistent.

Using the same experimental conditions, we further examine how stopping distance shapes pedestrians' interpretation of yielding intent. Under the same vehicle speed (high speed), the green curve in Fig.~\ref{fig4:TTA_behavior_validation}(b), left, rises earlier than the purple curve, indicating earlier crossing initiation when braking begins earlier and the vehicle stops farther from the crosswalk. The pedestrian model shows the same qualitative pattern in Fig.~\ref{fig4:TTA_behavior_validation}(b), right, suggesting that it captures the effect of stopping distance on yielding acceptance.

\subsubsection{Behaviors during crossing}

\begin{figure}[!t]
    \centering
    \includegraphics[width=\columnwidth]{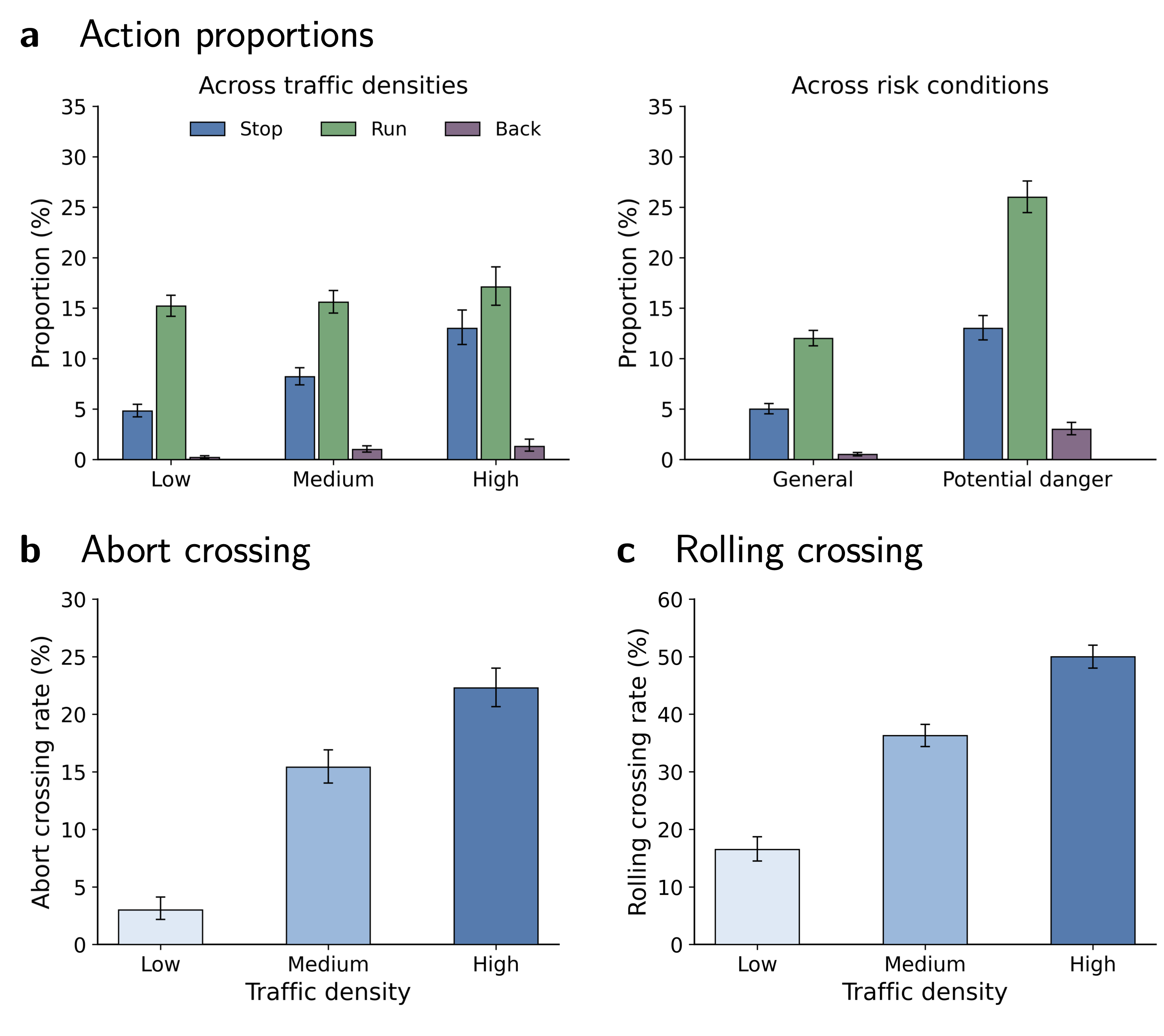}
    \caption{The model shows adaptive during-crossing behaviors under denser and riskier traffic.
    (a) Proportions of non-walk actions (\textit{stop}, \textit{run}, and \textit{back}) during crossing. The two subpanels show action proportions across traffic-density conditions and across context conditions, respectively. A time step was labeled as potential danger when the minimum TTA among observed approaching vehicles was below \(5\,\mathrm{s}\); otherwise, it was labeled as general. Non-walk actions increase with traffic density, and \textit{run}/\textit{back} actions are more frequent in potentially dangerous situations~\cite{almodfer2016quantitative}.
    (b) Proportion of episodes exhibiting abort crossing across traffic-density conditions. Abort crossing increases with traffic density~\cite{almodfer2016quantitative}.
    (c) Proportion of episodes exhibiting rolling crossing across traffic-density conditions. Rolling crossing increases with traffic density, indicating a shift toward staged crossing when traffic becomes denser~\cite{zhang2019evaluation}. Error bars denote 95\% confidence intervals.}
    \label{fig5:during_and_strategy}
\end{figure}

After roadway entry, the policy exhibited two forms of adaptive response: aborting an initiated crossing and using \texttt{run} or \texttt{back} actions in response to changing traffic risk. Fig.~\ref{fig5:during_and_strategy}(a) and (b) summarize how these behaviors varied across traffic-density and danger conditions.

\paragraph{Crossing abortions}
As shown in Fig.~\ref{fig5:during_and_strategy}(b), abort crossing occurs in a subset of episodes and becomes more frequent as traffic density increases. Binary logistic regression with traffic density coded as an ordered predictor confirms a significant increasing trend (OR per level \(=2.8\), 95\% CI [2.5, 3.0], \(p<0.001\)). This pattern is consistent with prior findings that retreat behaviors become more common under denser traffic~\cite{almodfer2016quantitative}, indicating that the model captures density-dependent abort crossing.

\paragraph{Run/back evasive movement.}
As shown in Fig.~\ref{fig5:during_and_strategy}(a), left, the use of \textit{run}/\textit{back} actions increases with traffic density. Trend tests confirm that both behaviors become more frequent under denser traffic: \textit{back} (OR \(=2.3\), 95\% CI [1.7, 3.2], \(p<0.001\)) and \textit{run} (OR \(=1.1\), 95\% CI [1.0, 1.2], \(p<0.1\)). 

We further examine their relationship with hazardous situations by defining approaching vehicles with TTA \(< 5\,\text{s}\) as a dangerous condition. As shown in Fig.~\ref{fig5:during_and_strategy}(a), right, the pedestrian model uses \textit{run} and \textit{back} more often in these situations, consistent with prior findings that pedestrians adopt more urgent evasive maneuvers under higher-risk traffic conditions~\cite{almodfer2016quantitative}.

\subsubsection{Crossing strategies}
For crossing strategies, we run randomized tests across different crosswalks and traffic-density conditions and compute the occurrence rate of rolling crossing. As shown in Fig.~\ref{fig5:during_and_strategy}(c), rolling crossings are relatively rare under low traffic density and become more frequent as traffic density increases. Binary logistic regression with traffic density coded as an ordered predictor (Low=1, Medium=2, High=3) confirms a significant positive trend (OR per density level \(=2.3\), 95\% CI [2.2, 2.5], \(p<0.001\)). Since one-stage and rolling crossings are complementary categories, this also implies a decrease in one-stage crossing with increasing traffic density. Overall, the results indicate that the model captures this adaptation of human crossing strategies, consistent with prior findings that rolling crossing becomes more prevalent under high traffic density~\cite{theofilatos2021cross}.

\subsection{Closed-Loop Outcomes and Baseline Comparison}

Pedestrian-vehicle collisions are rare but consequential events in real-world traffic~\cite{schneider2013pedestrian}. We therefore evaluate crossing outcomes and quantify the number of (1) successful crossing, (2) situations in which the agent decided not to cross, and (3) collisions. We, again, stratify these outcomes by the traffic density levels defined above.

\begin{figure}[!t]
    \centering
    \includegraphics[width=\columnwidth]{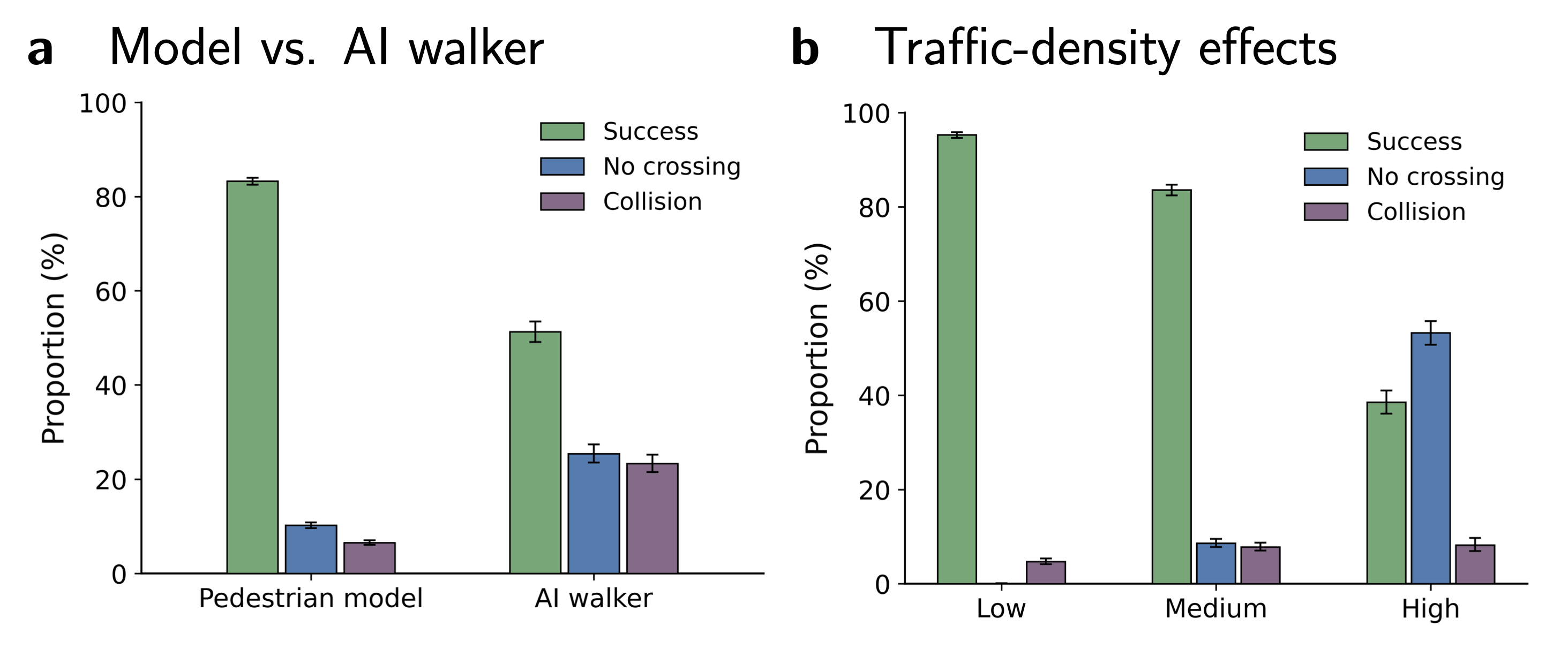}
    \caption{Closed-loop crossing outcomes and comparison with CARLA's AI walker.
    (a) Crossing outcomes for the proposed model and CARLA's AI walker under the same evaluation setting. The proposed model achieves a higher successful-crossing rate and lower no-crossing and collision rates.
    (b) Crossing outcomes across traffic-density conditions. The successful-crossing rate decreases with traffic density, whereas the no-crossing and collision rates increase~\cite{HesjevollElvik2016TrafficVolumeSafety}. Error bars denote 95\% confidence intervals.}
    \label{fig6:crossing_outcomes}
\end{figure}

\subsubsection{Overall and density-stratified outcomes}
Across 10{,}000 evaluation episodes, our pedestrian model achieves an overall success rate of 82.3\%, a no-crossing rate of 10.2\%, and a collision rate of 7.5\% (Fig.~\ref{fig6:crossing_outcomes}(a)). As shown in Fig.~\ref{fig6:crossing_outcomes}(b), increasing traffic density shifts outcomes toward lower success and higher no-crossing and collision rates. This trend is consistent with the finding that denser traffic reduces the availability of gaps and increases the risk of interaction~\cite{HesjevollElvik2016TrafficVolumeSafety}. Ordinal trend tests using logistic regression with traffic density coded as an ordered predictor (Low=1, Medium=2, High=3) confirm a strong increasing trend for no crossing (odds ratio per density level \(=15.2\), 95\% CI [13.3, 17.3], \(p<0.001\)) and a significant increasing trend for collisions (odds ratio \(=1.4\), 95\% CI [1.3, 1.5], \(p<0.001\)).

\subsubsection{Comparison with the AI walker}
As a baseline, we also evaluate CARLA's default \textit{AI walker} using the same environment and evaluation protocol across 2{,}000 simulation runs. As shown in Fig.~\ref{fig6:crossing_outcomes}(a), the AI walker exhibits a substantially higher collision rate (23.3\%) than the proposed model (7.5\%). This comparison shows that, while still higher than real-world collision rates, our pedestrian model is much closer to human data than the current default pedestrian models used in CARLA.

\subsection{Generalization and Adaptation}
We finally evaluate whether the learned policy transfers to new environments and can be efficiently adapted when traffic conditions change. We consider two forms of domain shift: transfer to an unseen map with broadly similar traffic density, and transfer to a traffic environment with different yielding norms and cultural settings. In both cases, we first evaluate the pretrained model zero-shot and then examine whether continued training improves adaptation.
\begin{figure}[!t]
    \centering
    \includegraphics[width=\columnwidth]{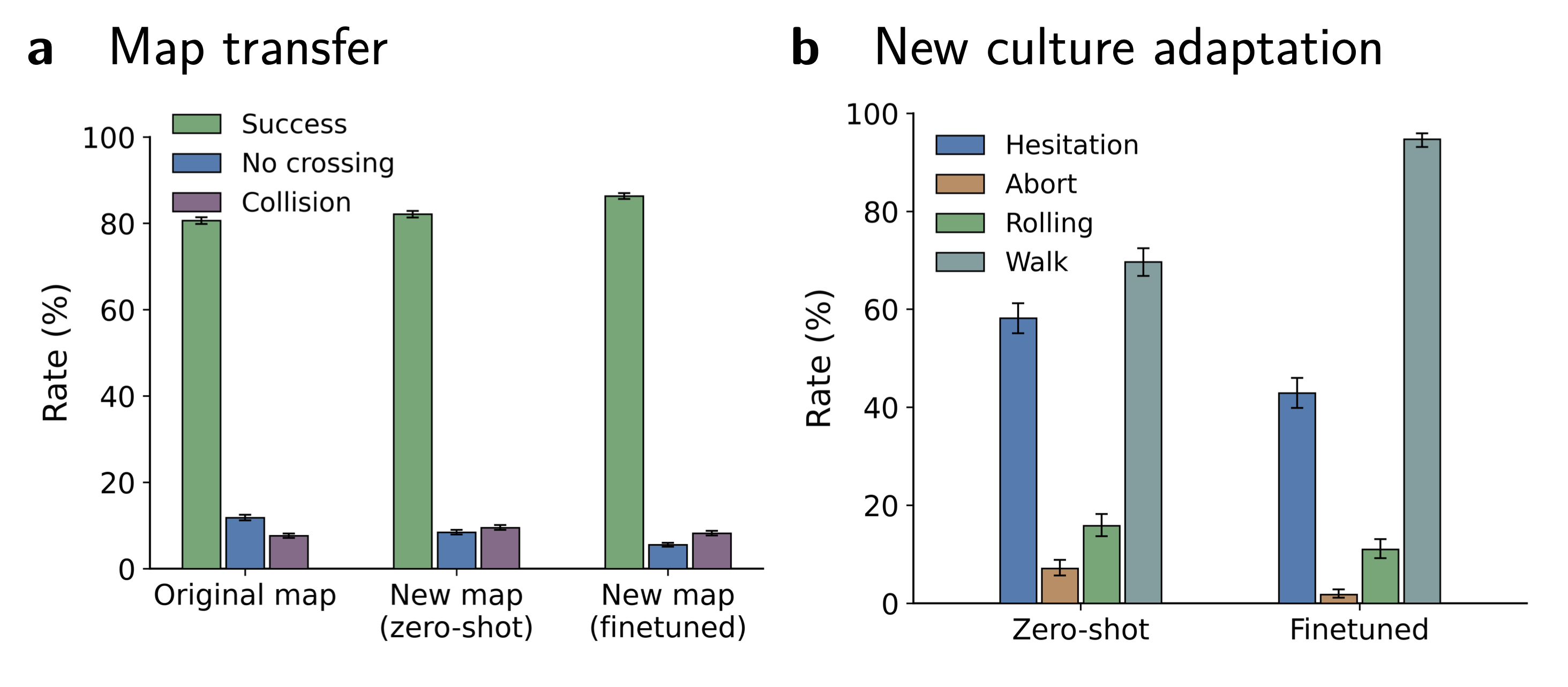}
    \caption{Generalization and adaptation across traffic environments.
    (a) Transfer to an unseen map under broadly similar traffic conditions. The pretrained policy retains comparable successful-crossing performance under zero-shot transfer, while fine-tuning further increases the successful-crossing rate and reduces no-crossing and collision rates relative to the zero-shot condition.
    (b) Adaptation to a target traffic regime with lower vehicle speeds, lower traffic density, and stronger vehicle yielding. After fine-tuning, the policy exhibits lower hesitation, abort-crossing, and rolling-crossing rates, less frequent \textit{run} and \textit{back} actions, and a higher proportion of \textit{walk} actions. Error bars denote 95\% confidence intervals.}
    \label{fig7:generalization_adaptation}
\end{figure}

\subsubsection{Transfer to an unseen map under similar traffic density}
We first examine generalization to a new map while keeping the overall traffic density broadly similar to that of the original domain. 
The results show that the model retains reasonable performance under this map transfer (Fig.~\ref{fig7:generalization_adaptation}(a)). In the original map, the model achieves a success rate of 82.3\%, a no-crossing rate of 10.2\%, and a collision rate of 7.5\%. When evaluated zero-shot on the unseen map, the success rate remains comparable at 82.1\%, while the no-crossing rate decreases to 8.4\% and the collision rate increases moderately to 9.5\%. These results indicate that the learned policy does not collapse under a change in map layout and preserves substantial crossing competence in a previously unseen environment, although safety degrades somewhat under zero-shot transfer.

We then continue training the pretrained policy on the new map. After finetuning, the success rate increases further to 86.3\%, the no-crossing rate decreases to 5.5\%, and the collision rate falls to 8.2\%. Compared with zero-shot transfer, this shows that a relatively small amount of additional training is sufficient to better align the policy with the new map and recover part of the lost safety margin. Overall, these results indicate that the pretrained policy retains substantial competence under map transfer and can be further improved through continued training in the target domain.

\subsubsection{Adaptation to different traffic norm and cultural settings}
Prior work suggests that pedestrian decision making is shaped by culture-dependent traffic norms and social context~\cite{nordfjaern2011cross,pele2017cultural}. We next examine a stronger form of domain shift in which the environment differs not only in map but also in interaction norms. Specifically, we construct a traffic setting with lower density, lower vehicle speeds, and substantially stronger yielding behavior, representing a context different from the original domain.

When evaluated zero-shot in this environment, the pretrained model already shows low collision risk, indicating robustness to easier and more cooperative traffic conditions. After continued training in the new domain, the policy shifts toward qualitatively different behavior (Fig.~\ref{fig7:generalization_adaptation}(b)). In particular, \textit{run} and \textit{back} become less frequent, while the pedestrian relies more on \textit{walk} and completes crossing in a smoother and more continuous manner.

Taken together, these two experiments show that the model retains nontrivial zero-shot performance in unseen environments and can be further adapted through continued training when the target environment differs in geometry or traffic norms.

\section{Discussion}

Our results lend evidence for formulating pedestrian crossing as a sequential decision-making problem under partial observability and uncertain vehicle behavior, showing an advance over prior work that has explored isolated gap-acceptance judgments, open-loop trajectory prediction, and manually scripted logic. Critically, behaviors such as hesitation, impatience, abort crossing, evasive speed adjustment, and density-dependent strategy selection need not be separately engineered. Instead, they emerge within policy when the model is equipped with human-like bounded perception, limited motor capabilities, and task incentives that reflect the trade-off between safety and efficiency.

The observed behavioral patterns are also consistent with the structure of the model. Because the pedestrian acts under noisy and partial observations of surrounding traffic, crossing decisions are made under uncertainty rather than from complete knowledge of the scene, which helps explain hesitation, tentative initiation, and retreat behavior. The reward design further imposes a unified trade-off between efficiency and risk: time pressure encourages eventual crossing, while collision and near-miss penalties discourage unsafe actions. This provides a common basis for the emergence of impatience, TTA- and speed-dependent gap acceptance, and sensitivity to yielding cues. In addition, the action space is important. Allowing the pedestrian to stop, walk, run, and move backward enables the policy to express aborts, evasive maneuvers, and rolling crossing as adaptive responses to traffic conditions. Traditional gap-acceptance or discrete-choice models~\cite{tian2022explaining,pekkanen2022variable,wang2025pedestrian} typically focus on crossing initiation only, whereas the present formulation models the full crossing process.

Rather than constructing a new pedestrian model for each system design or traffic regime, the results suggest that it is possible to train a reusable base policy and then adapt it through continued training to new environments. In our experiments, the pretrained model retained substantial competence on an unseen map and could be further aligned to new geometry and traffic conditions through fine-tuning. Adaptation to a more yielding and lower-speed traffic regime additionally showed that transfer should not be evaluated only in terms of collisions or success rates, but also in terms of changes in behavioral style. In the more yielding target traffic environment, adaptation was expressed less through dramatic changes in safety outcomes than through a shift toward smoother, more direct, and less evasive crossing behavior. This is important because the value of human-like pedestrian agents in simulation lies not only in whether they complete a crossing, but also in how they interact with vehicles while doing so.

Our findings have an important implication for the safety evaluation of automated vehicles and advanced driver-assistance systems. Safe automated mobility depends not only on detecting pedestrians, but also on anticipating how they respond to vehicle motion, yielding behavior, traffic density, and communication cues. Simulators that rely on fixed-path or scripted pedestrians may underestimate interaction risks by failing to expose vehicles to hesitation, tentative entry, retreat, and evasive speed changes. By providing closed-loop agents with such adaptive responses, the proposed approach may support more realistic testing of whether automated systems yield appropriately, communicate clearly, and remain robust across road layouts and traffic conditions.

At the same time, several limitations should be acknowledged. First, the present study focuses on crosswalk crossing in CARLA and does not yet cover a broader range of pedestrian situations such as jaywalking, signal compliance, multi-pedestrian interaction, or richer social negotiation with drivers and other road users. Second, although the model includes noisy perception and belief updating, its perceptual assumptions remain simplified relative to real human perception. It does not model richer occlusion reasoning, gaze behavior, intent inference, or semantic understanding of the scene. Third, the reward function remains designer-specified, which means that part of the resulting behavioral profile depends on assumptions chosen by the modeler rather than on direct estimation of latent human preferences from data. Fourth, comparison with human behavior in this paper is primarily at the level of empirical regularities and qualitative pattern alignment, rather than full calibration to large-scale human datasets or subject-level behavioral fits. Finally, our transfer results show that the model can generalize to new but related simulated settings, but they do not yet demonstrate broad generalization to real-world pedestrian behavior across diverse traffic situations.

These limitations suggest several directions for future work. Extending the model to signalized crossings, jaywalking, and multi-pedestrian interaction would broaden its ecological scope. Closer alignment with controlled behavioral experiments, immersive studies, or field data could strengthen empirical grounding and allow more rigorous calibration of model parameters and reward structure. Component-wise ablations could further clarify how perceptual uncertainty, decision timing, motor constraints, and individual reward terms contribute to the observed behavior. It would also be valuable to investigate whether variation in pedestrian age, mobility, risk preference, attention, or cultural background can be captured within the same framework, and whether some parts of the reward design can be inferred more directly from data.

\section{Conclusion}

This paper presented a theory-grounded pedestrian model for closed-loop automated-driving simulation. Pedestrian crossings are formulated as a partially observable sequential decision-making problem, and human-like behavior is generated by training a policy through reinforcement learning under perceptual, decision interval, and motor constraints. Our proposed model replicates human behavior consistent with ten empirically documented crossing phenomena including crossing initiation, adaptation during crossing, and multi-lane crossing strategy selection. In a closed-loop evaluation, the proposed model produces fewer collisions than CARLA's default AI walker under the same simulation conditions. Models pretrained in one specific traffic environment also retain realistic crossing performances in new unseen environments and can be readily fine-tuned to these new traffic environments. Overall, these results suggest that the pedestrian model proposed in this work provides a promising basis for simulator-ready pedestrian agents that combine behavioral realism, closed-loop deployability, and adaptability for automated-vehicle training, testing, and evaluation.

\section*{Conflict of Interest}

The authors declare no conflicts of interest.
\bibliographystyle{IEEEtran}
\bibliography{refs}

\end{document}